# Completeness of the Nearby White Dwarf Sample: Let Us Count the Ways

Terry D. Oswalt,[1] Jay Holberg,[2] and Edward Sion[3]

[1] *Embry-Riddle Aeronautical University, Daytona Beach, Florida, USA; terry.oswalt@erau.edu*

[2] *University of Arizona, Tucson, Arizona, USA; holberg@lpl.arizona.edu*

[3] *Villanova University, Villanova, Pennsylvania, USA; edward.sion@villanova.edu*

**Abstract.** We have recently extended our ongoing survey of the local white dwarf population, effectively doubling the sample volume. Based upon the latest distance determinations, Holberg et al. (2016) estimated the present 20 *pc* and 25 *pc* samples were about 86 and 68 percent complete, respectively. Here we examine how the completeness of the 25 *pc* sample depends upon other observables such as apparent magnitude, proper motion, photometric color index, etc. The results may provide additional clues to why "Sirius-Like systems" are underrepresented in the extended 25 *pc* sample and how additional nearby single white dwarf stars may be found.

## 1. The Nearby White Dwarf Sample

Only within the last decade or so has the sample of nearby white dwarfs become large enough and complete enough to provide a good basis for assessing their spatial distribution and demographics. Holberg, et al. (2008) used the *Catalog of Spectroscopically Identified White Dwarfs* (McCook & Sion 1999) to identify a sample of well over 100 white dwarfs likely to be within 20 *pc* of the Sun. Using available trigonometric and spectroscopic parallaxes, this sample was shown to be almost 80 percent complete.

Recently, Holberg, Oswalt & Sion (2016) expanded the search for nearby white dwarfs to a distance of 25 *pc*, expanding the original census by about a factor of two. Their initial analysis, based on trigonometric and spectroscopic parallaxes, indicated the 25 *pc* sample is about 68 percent complete. This expanded sample is large enough that its completeness also can be assessed via cumulative star counts by apparent magnitude and proper motion. In addition, it is complete enough that simple star counts can be used to derive the space density, avoiding the necessity of using the $1/V_{max}$ method or more sophisticated statistical methods to construct the white dwarf luminosity function, all of which are susceptible to small number fluctuations, observational biases, and/or unproven completeness — see García-Berro & Oswalt (2016) for a discussion of these problems.

## 2. Completeness as a Function of Distance

For a uniform space distribution of stars, cumulative counts should scale with the cube of the distance. Reliable trigonometric and/or photometric parallaxes are available for over 200 of the white dwarfs in the Holberg et al. (2016) 25 *pc* sample. Figure 1 displays in logarithmic form the cumulative counts vs. distance for this sample. Divergence of the least squares fit (red line) from the N-weighted line of slope = 3 (black line) in Figure 1 implies the sample is complete to about 13.4 *pc* and about 70 percent complete at 25 *pc*. The implied space density within the "complete" distance of 13.4 *pc* is 0.00463 $pc^{-3}$, in good agreement with Holberg et al. (2016).

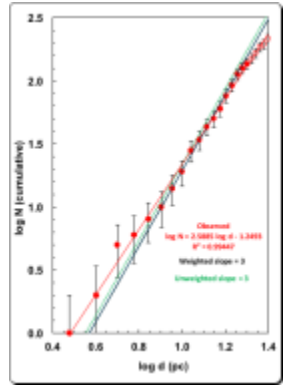

Figure 1. Cumulative counts in the 25-*pc* sample using best available distance estimates. Expected slope for a complete sample is +3. Error bars represent 1 / √N uncertainties. Red line denotes a least squares fit to the observed counts (only filled circles were used); green and black lines denote unweighted and N-weighted least squares fits to the expected slope = 3 for a complete sample. Inset: distance histogram.

## 3. Completeness as a Function of Proper Motion

Proper motions vary inversely with distance. Thus, cumulative star counts should scale with the inverse cube of proper motion. In general, proper motion data is more readily available than trigonometric parallaxes, as well as measured separately. This provides an independent way to assess the completeness of the "25 *pc* sample."

Figure 2 displays in logarithmic form the cumulative counts vs. proper motion for the 25 *pc* sample. Divergence of the least squares fit (red line) from the N-weighted line of slope = -3 (black line) in Figure 2 implies the sample is complete to about $\mu$ = 2.7 arcsec/yr. Assuming the space density derived from Figure 1, the average tangential velocity of the 25 *pc* sample is about 23 *km/sec*.

## 4. Completeness as a Function of Apparent Magnitude

For a uniform space distribution of stars, the inverse square dependence of apparent brightness on distance, the traditional logarithmic definition of apparent magnitude and the cubic dependence of volume on distance prescribe a cumulative star count that scales with apparent magnitude according to the relation (see Mihalas & Binney 1981):

$$logN(m) = 0.6m + constant$$

For nearby stars well within a Galactic scale height, such those in the 25 *pc* sample, a constant space density is a good assumption. Thus, deviation from the expected slope of 0.6 in the cumulative counts vs. apparent magnitude plot should be a good measure of the completeness. However, white dwarfs exhibit a range of absolute magnitudes and spectral energy distributions that span the filter bandpass, so the completeness is not necessarily constant across all apparent magnitudes (see Mihalas & Binney 1981). Assuming space density (*log n*) & absolute magnitude ($M_v$) don't depend on *V*, the expected cumulative counts should follow the relation:

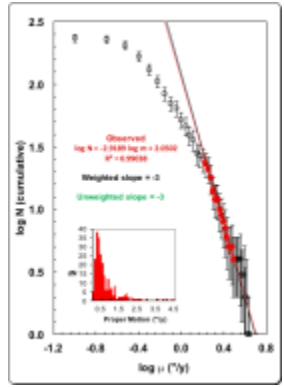

Figure 2. Cumulative counts in the 25-*pc* sample vs. proper motion. Expected slope for a complete sample is -3. Error bars and linear fits are as described in Figure 1. Inset: proper motion histogram.

$$logN = 0.6\,V - \log(3/4\pi) + \log n - 0.6 M_v$$

where *n* is the space density and $M_v$ is absolute magnitude. Figure 3 displays the cumulative counts in the 25 *pc* sample vs. *V* magnitude, corrected for the above effects. Deviation from the expected 0.6 slope of a complete sample occurs around $V = 15$.

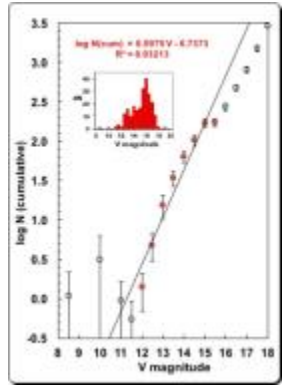

Figure 3. Cumulative counts in the 25-*pc* sample vs. *V* magnitude, corrected for the expected variation in space density with absolute magnitude described in the text. Expected slope for a complete sample is +0.6. Error bars and linear fits are as described in Figure 1. Inset: *V* magnitude histogram.

## 5. The White Dwarf Luminosity Function

Any observational study of the white dwarf luminosity function must be based on a well-defined sample. The 25 *pc* sample offers the most straightforward approach: simple star counts within a volume-limited sample that is demonstrably complete, or nearly so (for other approaches see the recent review by García-Berro & Oswalt 2016).

Figure 4 displays the white dwarf luminosity function for the 25-*pc* sample recomputed from the Holberg et al. (2016) data. Two stars in the 25-*pc* sample at $M_{bol} \sim 18$ and the more gradual slope beyond the downturn indicate an older age than the Harris et al. sample. The total space density implied by Figure 4 is $\sim 0.00319\ pc^{-3}$, implying the 25 *pc* sample is about 70 percent complete. Also note the "bump" at the bright end near $M_{bol} \sim 9$; Holberg et al. (2016) suggested this was caused by a two- to three-fold increase in the white dwarf birthrate during the most recent 0.5 *Gyr*.

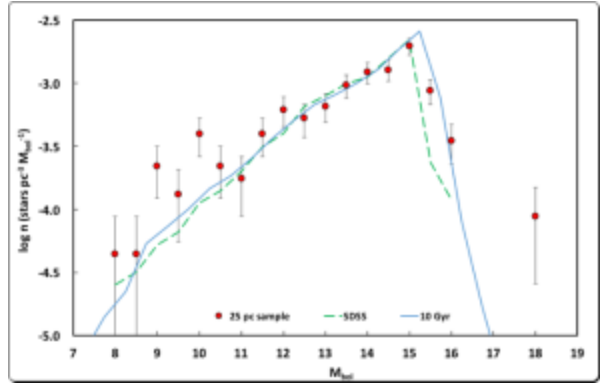

Figure 4. White dwarf luminosity function derived from the 25-*pc* sample (red circles). The white dwarf luminosity function derived from the Sloan Digital Sky Survey by Harris et al. (2006; green dashed line) and a model white dwarf luminosity function for a Galactic disk age of 10 *Gyr* computed by García-Berro (2016, priv. comm.; blue line) are shown for comparison. Error bars are as described in Figure 1.

## 6. Conclusions

Our analysis of other properties, such as reduced proper motions, space motions, color indices, spectroscopic and population subtypes, etc. of the 25-*pc* sample is ongoing. Over the next several years we will seek to construct a complete volume-limited sample of white dwarf stars to 50 *pc*. Early results from Gaia, particularly the high precision parallaxes, will greatly facilitate this effort.

**Acknowledgments.** We gratefully acknowledge funding for this project from NSF grants AST-1358787 to Embry-Riddle Aeronautical University, AST-1413537 to the University of Arizona and AST-1008845 to Villanova University. Special thanks to Enrique García-Berro for providing his most recently computed models. TDO also wishes to thank ERAU for travel support that enabled us to report these results at the EuroWD16 meeting.